\documentclass[12pt,preprint]{aastex}

\slugcomment{Submitted to ApJ}

\shorttitle{Metal abundances in NGC 6791}
\shortauthors{Villanova et al.}

\begin{document}


\title{NGC~6791: an exotic open cluster or the nucleus of a 
tidally disrupted galaxy?}


\author{Giovanni Carraro\altaffilmark{a,b}}
\affil{Astronomy Department, Yale University, New Haven, CT 06520$-$8101, USA}
\email{gcarraro@das.uchile.cl}

\author{Sandro Villanova}
\affil{Dipartimento di Astronomia, Universit\`a di Padova, Vicolo Osservatorio
2, I$-$35122, Padova, Italy}
\email{villanova@pd.astro.it}

\author{Pierre Demarque}
\affil{Astronomy Department,  Yale University, New Haven, CT 06520$-$8101, USA}
\email{demarque@astro.yale.edu}

\author{M. Virginia McSwain\altaffilmark{c}}
\affil{Astronomy Department,  Yale University, New Haven, CT 06520$-$8101, USA}
\email{mcswain@astro.yale.edu}

\author{Giampaolo Piotto}
\affil{Dipartimento di Astronomia, Universit\`a di Padova, Vicolo
Osservatorio 2, I$-$35122, Padova, Italy}
\email{piotto@pd.astro.it}

\and
\author{Luigi R. Bedin}
\affil{ESO, K. Schwarzschild Str. 2, 85748 Garching, Germany }
\email{lbedin@eso.org}


\altaffiltext{a}{Departamento de Astr\'onomia, Universidad de Chile,
    Casilla 36-D, Santiago de Chile, Chile} 
\altaffiltext{b}{Andes
    Fellow, on leave from Dipartimento di Astronomia, Universit\`a di
   Padova, Vicolo Osservatorio 2, I$-$35122, Padova, Italy}
\altaffiltext{c}{NSF Astronomy and Astrophysic Postdoctoral Fellow}

\begin{abstract}
We report on high resolution Echelle spectroscopy of 20 giant stars in
the Galactic old open clusters NGC 6791 obtained with Hydra 
at the WIYN telescope. High precision radial velocity allow us to
isolate
15 {\it bona fide} cluster members. From 10 of them we derive a 
global [M/H]=+0.39$\pm$0.05. We therefore confirm that NGC 6791
is extremely metal rich, exhibits  a few marginally sub-solar 
abundance ratios, and within the resolution of our spectra does not
show evidences of spread in metal abundance. With these new data we re-derive
the cluster fundamental parameters suggesting that it is about 8 Gyr
old and 4.3 kpc far from the Sun. 
The combination of its chemical properties, age, position, and Galactic
orbit hardly makes NGC 6791 a genuine Population I open cluster.
We discuss possible interpretations of the cluster peculiarities
suggesting  that the cluster might be what remains
of a much larger system, whose initial potential well could have
been sufficient to produce high metallicity stars, and which
has been depopulated by
the tidal field of the Galaxy. Alternatively, its current
properties may be explained by the perturbation of the Galactic bar
on an object originated well inside the solar ring,
where the metal enrichment had been very fast.
 
\end{abstract}

\keywords{open clusters: general --- open clusters: (\objectname{NGC 6791})}

\section{Introduction}
NGC~6791 is an extremely interesting and intriguing open cluster.  The
combination of old age, small distance and high metal abundance makes
this cluster very attractive, and indeed in the last 40 years it has been
the target of intensive and numerous studies (Carney et al. 2005, and
references therein).  A large number of optical photometric studies
(Stetson et al. 2003 and references therein) has been recently
complemented by
the deep ACS/HST investigation by King et al. (2005), and
the near IR study by  Carney et al. (2005).\\
Since the pioneering study of Kinman (1965) it was clear that
NGC 6791 is a very old and very metal rich cluster.

Its age was measured several times by using different sets of isochrones
(Carraro et al. 1999 and references therein), and it is probably confined
in the range from 8 to 12 Gyr, depending on the cluster precise
metal abundance. Taylor (2001, and references therein) critically reviewed
all the available metallicity estimates, concluding that the [Fe/H] for NGC 6791
should probably lie in the range +0.16 to +0.44 dex.
This combination of age and metallicity is unique in the Milky Way
open cluster population, and has been recently questioned by Bedin et al.
(2005), whose HST study of the White Dwarf  cooling sequence supports
a much younger age. As these authors comment, this age discrepancy
may arise from defects in the White Dwarf current models, or  from the poorly
known cluster metal abundance.\\

Noteworthy, the cluster is also known to harbour a number of sdB/sdO stars 
(Landsman et al. 1998, Buson et al. 2005), which may be
explained by a  scenario
of a high metallicity driven wind in the Red Giant Branch phase of the
progenitors of these stars or, more simply, by the binarity 
hypothesis (Green et al. 2005) .
 
The UV upturn (namely the abrupt rise in the UV continuum emission
shortward of $\lambda \approx 2000\AA$)  similar to that typical of any ellipical galaxy (Landsman et al.
1998) and the highly eccentric orbit, unusual for a Population I object, 
contribute to make this cluster even more intriguing .\\
In an attempt to substantially improve our knowledge of NGC ~6791, 
we carried 
out a spectroscopic campaign to provide radial velocities 
and accurate metallicities of a statistically significant number of stars
in  the cluster. In fact, current abundance determinations either lack
sufficient resolution or are restricted to a very small number of stars.\\
This new set of abundance estimates coupled with the high quality of 
existing photometry (Stetson et al 2003) allow us to significantly
improve on the fundametal parameters of this cluster, and 
better clarify its intriguing nature.

\section{Observations}
The observations were carried out on the night of July 28,
2005 with the Hydra spectrograph 
at the WIYN telescope at Kitt Peak observatory under photometric conditions and
typical seeing of 1\farcs1 arcsec. 
The MOS consists of the Hydra positioner, which in 20 minutes 
can place 89 fibers within the 1-degree diameter focal plane of the
telescope to $\approx$ 0.2 arcsec precision. This project employed the 3-arcsec 
diameter red-optimized fiber bundle. The fibers feed a bench-mounted 
spectrograph in a thermally isolated room. With the echelle grating and
Bench Spectrograph Camera the system produces a resolution of 20,000 at 
6000 \AA~. The wavelength coverage of 200 \AA~ around the central
wavelength of 
6000 \AA~ provides a rich array of narrow absorption lines.
We observed 20 RGB/clump stars with 45 min exposures,
for a grand total of 4.5 hours of actual photon collection time 
on the same single star.
The 20 stars where selected from the Stetson et al. (2003)
photometric catalog to be giant stars and to have the
right magnitudes to be observed with the WIYN 3.6m telecope.
We  restricted the sample to giant stars brighter than
V $\approx$ 15.
The stars are listed in Table~1, where first column reports
Stetson et al. (2003) numbering and  column 2 Kinman (1965) numbering
(K65).
Then coordinates, magnitudes and colors are taken
from Stetson et al. (2003).
The radial velocities and spectral classification
have been derived in this paper, following Villanova et al. (2004).

\section{Data Reduction}
Images were reduced using IRAF\footnote[2]{IRAF is distributed by the
National Optical Astronomy Observatories, which are operated by the
Association of Universities for Research in Astronomy, Inc., under
cooperative agreement with the National Science Foundation.},
including bias subtraction, flat-field correction, frame combination,
extraction of spectral orders, wavelength calibration, sky subtraction
and spectral rectification. The single orders were merged into a
single spectrum. As an example, we show in Figure~2 a portion of the
reduced, normalized spectrum for 
star $\#$11814 where some spectral lines are
identified. Some spectra have very low S/N, although all the observed
stars have practically the same magnitude.  This could happen for two
reasons: the first one is an imperfect pointing of the fiber and the
second one a possible bad fiber-trasmission. Because of this, we 
could not use five
stars for abundance measurements (see below).

\section{Radial Velocities}
Radial velocities (RV) for RGB and clump stars in NGC~6791 have been
determined many times in the past. Kinman (1965) obtained radial velocities
for 19 stars, and spectral type for 21. Later, RVs
have been measured by Geisler (1988, 12 stars), Friel  et al. (1989, 9 stars), 
Garnavich et al. (1994, 18 stars), Scott et al. (1995, 32 stars) 
and Friel et al. (2002, 41 stars), with different resolution
and precision.

\noindent
We derived here RVs for 20 stars (see Table~1).  The
radial velocities of the target stars were measured using the IRAF
FXCOR task, which cross-correlates the object spectrum with 
a template. 
As template, we used a synthetic spectrum calculated by 
SPECTRUM (see 5.2 for a description of the program) with roughly
the same atmospheric  parameters and metallicity of the observed
stars.  The final error in the radial velocities was typically less
than 0.2 km s$^{-1}$, and in many cases less than 0.1 km s$^{-1}$.
These errors are significantly lower than in any other previous
investigation.  This allowed us to clean out field interlopers and
isolate 15 {\it bona fide} members.  RVs are plotted in
Figure~3.  Five stars have radial velocites completely different from
the others, and so were considered non-members, although it possible
that some of them are binary stars.\\ 
A few of our targets are in common with previous investigations, and
we can have an external check on our RV measurements.
For 11 stars we provide the
first estimate of the RV. \\ 
In general, we find that Garnavich
et al. (1994) RV estimates (for stars 3003, 3010, 3036, 2001, 3018 and 2008)
are systematically larger than ours, by about 5-7 km s$^{-1}$, although,
given their typical large error (5 to 15 km/sec), 
these differences cannot be considered statistically significant.
Also Kinman (1965) RVs for the two stars in common with our
investigation (stars 2001 and 2008) are larger than our estimate.  
The largest deviation is with respect to the RVs
by Friel et al. (2002), and their previous
measurements (Scott et al. 1995, Friel et
al. 1989).  In this case, there are differences in some case exceeding 
20 km s$^{-1}$ (stars 3003 and 2008).  Finally, within the errors, we
find a good agreement for the two stars (stars 3003 and 3010) we have in
common with Geisler (1988).\\ 
From the RVs of the 15 cluster members in our sample, we obtain a mean
radial velocity $V_r=-47.1\pm0.8$ km s$^{-1}$, in good agreement,
within the errors, with the values obtained by the other authors, with
the exception of Kinman (1965). The RV dispersion results to be 
$\sigma_r=2.2\pm0.4$ km s$^{-1}$.

\section{ABUNDANCE ANALYSIS}

\subsection{Atmospheric parameters}
The atmospheric parameters were obtained from the photometric $BVI_C$
data of Stetson et al.  
(2003).
According to 
Stetson et al., the 
most likely reddening and absolute distance modulus are E(B-V ) = 0.09
[(E(V-I) = 0.11)] and $(m-M)_0=12.79$.  Effective temperatures 
($T_{\mathrm eff}$ )
were
obtained from the color-$T_{\mathrm eff}$ relations of Alonso et al. (1999),
Sekiguchi \& Fukugita (2000), and Ramirez \& Melendez (2005).  
The temperatures, obtained from the B-V and V-I colors using the quoted
relations, are in agreement within 50-100 $^0K$.  The gravity log(g) was
derived from the canonical formula :\\ $log(\frac{g}{g_{\odot}}) = 4
\times log(\frac{T_{\mathrm eff}}{T_{\odot}}) -
log(\frac{L}{L_{\odot}})+ log(\frac{M}{M_{\odot}})$ \\ In this
equation, the mass $M/M_{\odot}$ was derived from Straizys \&
Kuriliene (1981).  The luminosity $L/L_{\odot}$ was derived from the
absolute magnitude $M_V$, adopting the distance modulus of Stetson et
al. (2003).  The bolometric correction (BC) was derived from the
BC-$T_{\mathrm eff}$ relation from Alonso et al. (1999).  The typical
error in log(g) is 0.1 dex.\\ Finally, the
adopted
microturbulence velocity is the mean of the values given by the relation
(Gratton et al. 1996): \\ $v_t = (1.19 \times 10^{-3}) T_{\mathrm eff} -0.90
log(g) -2$; \\ and the relation (Houdashelt et al. 2000): \\ $v_t = 2.22
- 0.322 log(g)$. \\ The typical error in $v_t$ is 0.1 $km s^{-1}$.

\subsection{Abundance determination}
The resolution (R=17,000 at  6580\AA) of our spectra, the high metallicity of
the cluster, and the low temperature of the target stars cause a lot
of blending, and therefore it was not possible to measure the
equivalent width of the single spectral lines. For this reason, the
abundances were determinated by comparing the observed spectra with
synthetic ones. The synthetic spectra were calculated by running
SPECTRUM, the Local Thermodynamical Equilibrium (LTE) spectral synthesis program freely distributed by
Richard O. Gray (see Piotto et al. 2005 for the details on our
synthetic spectra calculation). Model atmospheres were interpolated
from the grid of Kurucz (1992) models by using the values of
$T_{\mathrm eff}$ and log(g) determined as explained in Section 5.1. We
analized stars with $T_{\mathrm eff} > 3900 ^0K$ because, for lower
temperatures, molecular bands were present, creating difficulties for
continuum determination. We could analyze only ten stars, 
after rejecting the non-members and the stars which resulted to be
too faint and too cool.

First of all, we compared the entire spectrum (range 6400-6760 \AA) with
the synthethic one in order to obtain an estimate od the global
metallicity [M/H]. Then, we analized single lines in order to measure
abundances of Fe, Ca, Ti, Ba, Al, Ni, and Si.

A preliminary lines-list was obtained 
considering all the strongest lines present in our
spectra, identified using the line-list distributed
with SPECTRUM.  The final
lines-list was created from the preliminary one by comparing the observed
solar spectrum with a sinthetic one, calculated  with SPECTRUM
for the Sun
parameters ($T_{eff} = 5777 K, log(g) = 4.44, v_t = 0.8 m/s$).
The lines in the synthetic spectrum which did not properly match the
observed ones were rejected.  
We also checked whether the preliminary line identification
was correct using the MOORE line database (Moore et al. 1966).
Table~3 shows the final list of lines we used in our analysis. The
resulting metallicities for each star are listed in Table~4.

Due to the radial velocity shit, a few lines (6462.567, 
6475.630, 6493.781,  6532.890 \AA) overlapped with telluric lines.
We did not consider these lines in the abundances determination.

An example of the comparison between synthetic and observed spectra is
shown in Fig.~4, for the case of star $\#$11814 Finally, using the stellar
parameters [colors, $T_{\mathrm eff}$, and log(g)] and the absolute
calibration of the MK system (Straizys \& Kuriliene 1981), for each
star, we derived the stellar spectral classification (see Villanova et al. 2004
for details), which is listed
in Col.~9 of Table~1.\\

\noindent
The weighted mean of the [Fe/H] content of the ten members
of NGC~6791
analysed in the present paper is [Fe/H] = +0.39$\pm$0.01
(internal error).
Previous investigations reported a variety of estimates for the
metallicity of NGC 6791.  By using low resolution spectroscopy, Friel
\& Janes (1993) obtained [Fe/H] = +0.19$\pm$0.19 from 9 stars, and
Friel et al. (2002) [Fe/H] = +0.11$\pm$0.10 from moderate resolution
spectra of 39 stars.  Because of the large errors, the first estimate
is compatible with our one, within one sigma, while 
the second one is off by almost $3\sigma$.\\

\noindent
Interestingly enough, our results are in very good agreement with the
study by Peterson \& Green (1998), who derived [Fe/H] = +0.40$\pm$0.10
for star $\#$2017, a cool blue horizontal branch star, using a resolution very similar to the one used in
the present study.\\

\noindent 
In conclusion, our results confirm that NGC 6791 is actually a very
metal rich cluster. Noteworthy, within the errors of our measurements,
the metallicities listed in Table~4 do not show any significant
abundance spread.

\subsection{Abundance ratios}
Abundance ratios constitute a powerful tool to assign a cluster to a
stellar population (\citealt{fri03}, \citealt{car04},  \citealt{vil05}).\\
In Table~5 we list the abundance ratios for the observed stars in
NGC~6791. These values do not show any particular anomaly.
All the abundance ratios are solar scaled,
with the only exception of [Al/Fe] and [Ba/Fe], which seem to be
slighlty under-abundant. Our ratios are in
good agreement with those provided by Peterson \& Green (1998).

\section{Distance and age of NGC 6791}
Our accurate determination of the metal content of NGC 6791 
allows a new, more reliable estimate of the cluster 
distance and age.\\
To this purpose, in this section, we are going to fit the observed
Color Magnitude Diagram (CMD) from Stetson et al. (2003) with both the Padova (Girardi et al. 2000) and
Yale-Yonsei isochrones (Yi et al. 2001; Demarque et al. 2003).
Previous similar studies allowed to confine the cluster age in the
relatively large interval between 8 and 12 Gyrs (Carraro et al. 1994,
Chaboyer et al. 1999,
King et al. 2005, Carney et al. 2005, and reference therein).  As
discussed in Stetson et al (2003), and confirmed in this work,
the cluster reddening is E(B-V) =
0.09$\pm$0.04. There is a large scatter in the literature on the
absolute distance modulus $(m-M)_o$ estimates, which range in the
interval between 12.6 and 13.6.  These large uncertainties in both age
and distance have been usually ascribed to uncertainties in the
cluster metal abundance. The new specroscopic data presented in this
paper allow to put on a more solid basis these fundamental parameters.

\subsection{Padova Isochrones}
Our [Fe/H] = 0.39 empirical measurement, translates into a metallicity
Z = 0.046 (Carraro et al. 1999)
and implies a $\Delta Y/\Delta Z$ close to 2.
We generated isochrones for this metallicity, and for ages ranging from 7
to 11 Gyrs from Girardi et al. (2000).  
An appropriate and meaningful isochrone fit implies that all the loci
of the CMD, e.g. the turn-off point (TO), the sub giant branch (SGB),
the red giant branch (RGB), and the clump of He-burning stars must be simultaneously 
overlapped by the models. Our best fit (by eye) estimate is shown in
Fig.~5 and 6, both in the V vs (B-V) and V vs (V-I) plane.  In Fig.~5
we plot the V vs (B-V) CMD of NGC~6791, and superpose the whole set of
isochrones, whereas, in Fig.~6, we only show the best fit isochrone in
the V vs (B-V) and V vs (V-I) plane.

The isochrone solutions in Fig.~5 have been obtained by shifting the
theoretical lines by $E(B-V)$ = 0.09 and $(m-M)_V$ = 13.35.  Clearly ages
older than 9 Gyr can be ruled out, since a fit to the TO with
an older isochrones implies to decrease the distance modulus, but, in
this way, the theoretical clump would be brighter than the observed
one.  On the other hand, also ages younger than 8 Gyr do not seem
possible, since a fit of the TO region with a younger
isochrone would result in a RGB redder than the observed one (impling
a reddening value significantly larger that the observational limits),
and also the clump magnitude would be fainter than the observed
counterpart.  

Only the isochrones for ages of 8 and 9 Gyr provide a good fit.
In details, the 9 Gyr isochrone fits well the CMD with the adopted
parameters, although the clump luminosity turns out to be slightly
brighter than the observed one.
On the other hand, the 8 Gyr isochrone must be shifted by $E(B-V)$ = 0.09
and $(m-M)_V$ = 13.45 to provide a very good fit.
This is shown in Fig.~6. We note that the lower MS is mismatched.
This is a well known problem for metal rich clusters, as extensively
discussed by Bedin et al. (2001), and it is likely due to problems in
the transformation of the models from the theoretical to the
observational plane.

On the overall, however, the fit is very good, and implies for NGC 6791
this set of fundamental parameters: 
8.0$\pm$1.0 Gyr, 
13.07$\pm$0.05 (internal error), 
and 0.09$\pm$0.01
for the age, 
absolute distance modulus, and reddening, respectively.
The associated errors 
are internal errors and have been estimated by eye. They simply reflect
the degree of freedom we have to displace the isochrones, still
achieving an acceptable fit.

\subsection{Yale-Yonsei Isochrones}
An independent determination of the age, distance, and a constraint on
the reddening of NGC~6791 can be derived using the $Y^2$ isochrones
(Yi et al. 2001; Demarque et al. 2003).  The Padova and $Y^2$
isochrones were both constructed using the same OPAL opacities tables.
Otherwise, the description of the microscopic and macroscopic physics,
as well as the numerical procedures, differ in many details in the two
sets of isochrones.  The color transformations are also independently
derived.

In the $Y^2$ system, in which $Z_{\odot}$ = 0.0181, 
[Fe/H] = 0.39 corresponds to $(Y, Z)$ = (0.31, 0.04).  The fit
is based on the main sequence (MS) position just below the TO, the position
of the TO point, the SGB, and the RGB color. A good fit, as shown in
Fig.~7, is obtained for $(m-M)_V$ = 13.35, 0.1 mag smaller than the  distance modulus we
used for the Padova isochrone fit.  The best fit is obtained by
assuming a reddening $E(B - V)$ = 0.13, somewhat larger than the
reddening adopted in the previous section, but still within the
estimated range 
of previous investigations.
The age we derive from the $Y^2$ fit is between 8 and 9 Gyr,
in good agreement with the Padova age, even though one notes that the
position of the lower main sequence differs in the two sets of
isochrones.  The unevolved main sequence of the Padova isochrones has
a steeper downward slope than the observations, whereas the opposite
holds for the $Y^2$ isochrones.

Similarly, Chaboyer et al. (1999) conclude that the cluster age is 8.0
$\pm$ 0.5 Gyr, assuming [Fe/H] = 0.4, but using an older observational data
set (Kaluzny \& Rucinski 1995),
and a version of the Yale stellar evolution code that slightly
differs from the one used to construct the $Y^2$ isochrones.  In an
analysis of their infrared photometry, Carney et al. (2005) derive an
age between 9 Gyr (for [Fe/H] = 0.3) and 7.5 Gyr (for [Fe/H] = 0.5),
also in good agreement with our result.  Both the Chaboyer et
al. (1999) and the Carney et al. (2005) ages are consistent with the
Padova and $Y^2$ fits described in this paper.  We should note however
that the Carney age estimates were obtained using the same set of
$Y^2$ isochrones that we used in the present work, and therefore
their age determination is not completely independent form our one.

Stetson et al. (2003) derive a much older age (12 Gyr) with the help
of unpublished VandenBerg isochrones.  It appears that the large
difference in age is due in part to the authors' choice of a markedly
smaller absolute distance modulus (12.79 mag).  A superposition of the $Y^2$
isochrones for the range 8-12 Gyr, shifted by $(m-M)_0$ = 12.79 and $E(B
- V)$ = 0.09 (Stetson et al. adopted values), on the CMD of NGC 6791
is shown in Fig.~8, for comparison purpose.  Although it is not
possible to rule out completely the Stetson et al. (2003) fit, the
disagreement with other well calibrated isochrones raises questions
about the calibration of the VandenBerg isochrones.

An additional, independent age and distance estimate is in King et
al. (2005) who obtained an excellent fit of the upper main sequence,
TO, and SGB both of Stetson et al. (2003) groundbased CMD of NGC 6791,
and of their ACS/HST CMD in the F606W, F814W bands by using the Teramo
isochrone set by Pietrinferni et al. (2004). From both fits, King et
al. (2005) derived an age of $9\pm1$ Gyr, an absolute distance modulus
$(m-M)_0=13.0$, and a reddening E(B-V)=0.12 for a metallicity
[M/H]=+0.4, and Y=0.288. Also in the fit of the $m_{\mathrm F814W}$
vs. $m_{\mathrm F606W}$-$m_{\mathrm F814W}$ ACS/HST CMD, the Teramo
isochrones tends to be redder and redder going to fainter magnitudes,
starting from $\sim2$ magnitudes below the TO, as already noticed for
the Padova isochrones.

\noindent
Finally, we must mention that the precise age of NGC~6791 also depends
on the adopted value of the ratio $(\Delta Y/\Delta Z)$ for galactic
helium enrichment.  This quantity is poorly known; it may be a
function of Z, and may differ from system to system.  Demarque et al.
(1992) have found that varying Y from 0.32 to 0.36 could reduce the
age of the cluster by as much as 15\%.  The age estimate of NGC~6791
might have to be increased if the enrichment ratio $\Delta Y/\Delta Z$
is much less than 2 (isochrones for $\Delta Y/\Delta Z$ near 2 were
assumed in the Padova, $Y^2$, Teramo, and VandenBerg fits).

With our present 
knowledge of the $\Delta Y/\Delta Z$ parameter, we conclude that 
the age of NGC 6791 must be in of
range 7.5-8.5 Gyr,
with a higher preference towards the higher limits.
We point out that, even in the unlikely event that the age of
NGC~6791 is as low as 7.5 Gyr, its high metallicity 
and age
present a major
challenge to the accepted view of Galactic chemical enrichment.  \\

\noindent
In conclusion, 
three sets of independent isochrones consistently inply that the age of NGC 6791 
is around 8 Gyr, adopting the 
metallicity and the reddening coming from observations. The difference
in reddening might simply ascribed to the different Helium abundance
adopted,
and to some photometric zero point error.

\section{DISCUSSION and CONCLUSIONS}
The hardest point with NGC~6791 is how inside this cluster such high
metallicity stars could have been
produced. In fact, this cluster does not
have any counterpart in the Milky Way.  We note here that Kinman
(1965) originally identified NGC~6791 as a globular cluster.  Even if
this interpretation were to be adopted, the high metallicity of
NGC~6791, much higher than that of any Galactic globular cluster or
nearby dwarf galaxy, remains mysterious. \\

With Berkeley~17 and Collinder~261, NGC 6791 is one of the oldest open
clusters of the Galaxy (Carraro et al. 1999), but its metal abundance is incomparably
higher. Moroever, NGC 6791 is one of the most massive open clusters
(4000 $M_{\odot}$ at least). It lies at 1 kpc above the Galactic plane,
inside the solar ring. This combination of mass and position is hard
to explain, since the interaction with the dense Galactic environment
should strongly depopulate a typical open cluster. \\

NGC 6791 is routinely considered in the 
studies of the
chemical evolution of the
Galactic disk, and occupies a unique position in the Galactic disk
radial abundance gradient (see Fig.~9).  By including NGC 6791, the
slope of the gradient changes from -0.05 (solid line) to -0.07 (dashed
line).  Besides, if one considers the slope defined only by clusters
older than 4 Gyr (Friel et al. 2002, Fig.~3, upper panel), the slope
doubles, from -0.06 to -0.11.\\ In the same figure, the horizontal solid
line indicates the epicyclical amplitude of NGC 6791 orbit (see below,
and Carraro \& Chiosi 1994).  One can readily see how NGC 6791 is
quite an exotic object.  
If, by chance, at the present time the cluster would be at different
orbit phase, which would put it, e.g., beyond 12 kpc, 
there would be a drastic change and even
an inversion of the slope of the 
Galactic disk abundance gradient.  Finally, the position of this
cluster in the Galactic disk Age-Metallicity relationship (Carraro et
al. 1998) is puzzling as well, since the cluster significantly deviates
from the mean trend.\\

In Fig~10 we present NGC~6791 Galactic orbit. This was obtained
by integrating back in time ( 1 Gyr) the cluster from its present 
position and kinematics
using the Galaxy N-body/gasdynamical model by Fux (1997, 1999).
The adopted radial velocity and proper motions
come from Geisler (1988) and Cudworth (private comunication),
whereas the Galactocentric rectangular  initial conditions (positions and velocities)
was derived as in Carraro \& Chiosi (1994).\\
\noindent
Intererstingly enough, this plot shows that the cluster moves 
from the outer disk regions of the Milky Way, more than 20 kpc far away from the Galactic center,
and enters the Solar Ring 
going as close as 6 kpc from the Galactic center.\\

The eccentricity (e=0.59) of this orbit is quite high for a Population I
star cluster (Carraro \& Chiosi 1994), and it is much 
more
similar to a globular cluster/dwarf galaxy orbit.

A plausible scenario  is that NGC 6791 is what remains
(the nucleus) of a much larger system, which underwent strong tidal
disruption. This would explain the cluster orbit, and provide
a reasonable explanation for the high metallicity of its stars,
which  could have been produced only inside a deep potential well.

However, within the observational errors, we did not find
any significant abundance spread. This would mean that the bulk
of the stars in the cluster was produced in a single burst
of star formation. This fact makes more difficult
the capture interpretation since Local Group Galaxies
normally exhibit spreads in metal content and possess
lower metal abundance (Mateo 1998).
We stress however the fact that
our results are based on only ten stars, and that only larger
spectroscopic surveys can better address this particular
problem.\\

An alternative more conservative scenario is that the cluster was born
in the inner side of the Galaxy, close to the bulge, where the metal
enrichment has been  fast. 
Grenon (1999) studied the kinematics
of a group of old (10 Gyrs) metal rich $[M/H] \geq 0.30$ 
stars and suggested that they formed close to the bulge and then migrated at large
Galactocentric distance due to the perturbation of the Galactic bar.

The orbit we calculated actually includs the effect of the bar, and NGC 6791
indeed moves well outside the solar circle.
NGC 6791 is very concentrated for an open cluster, and spent most of its time
at moderate Galactic latitude. This might help to explain
its survival.

\acknowledgments The observations described in this paper were carried
out remotely from Yale University by Giovanni Carraro. We deeply
acknowledge Diane Harmer, George Will and Chris Hunter for support
and help.  
The work of GC is
supported by {\it Fundacion Andes}. 
GP and SV acknowledge the support
by the italian MIUR, under the program PRIN2003.
MVM is supported by an NSF Astronomy and Astrophysics
Postdoctoral Fellowship under award   AST$-$04011460.
PD's research is supported in part by NASA grant
NAG5-13299.

\clearpage

\clearpage

\begin{figure}
\includegraphics[scale=0.8]{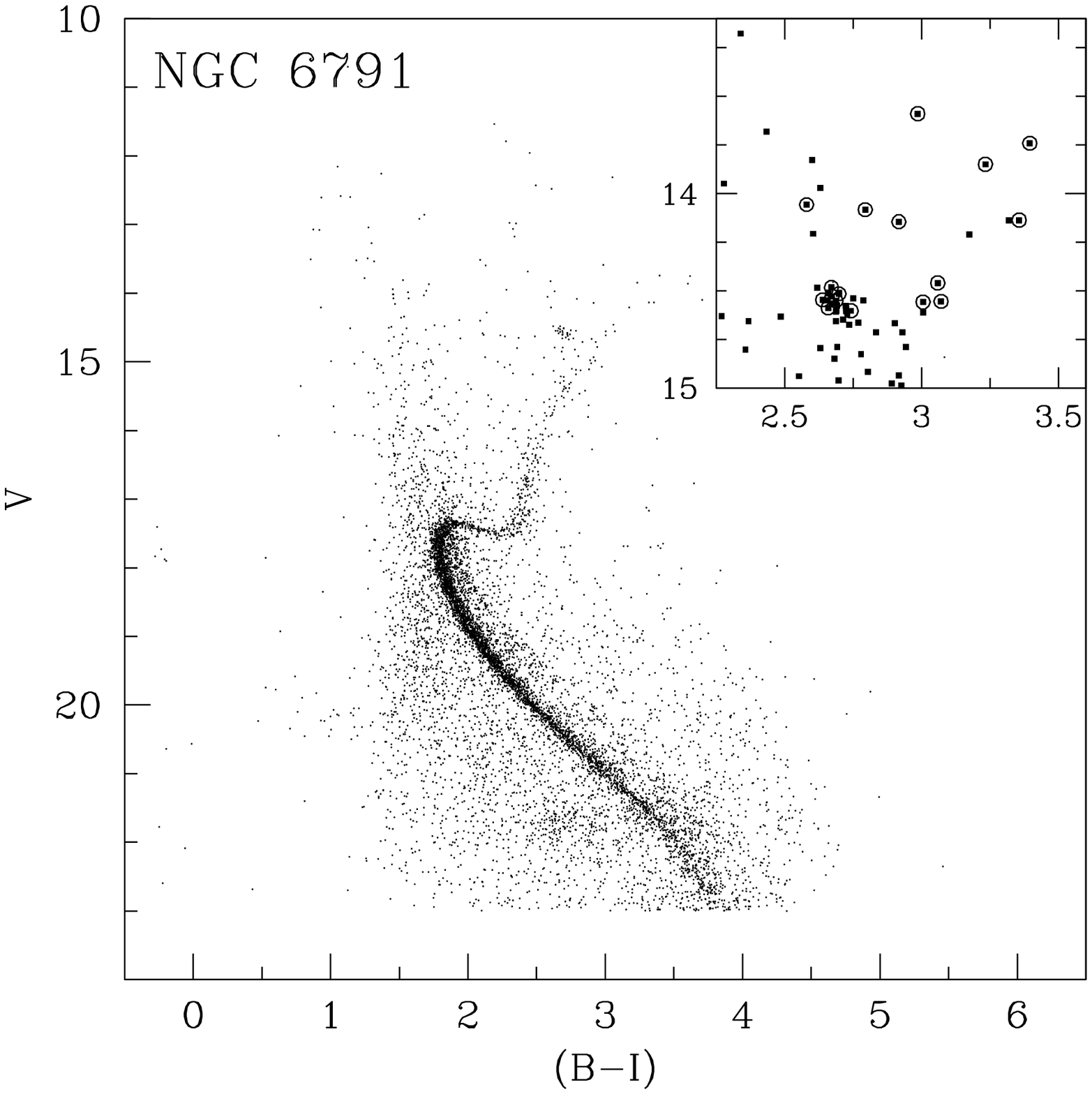}
\caption{The CMD of NGC 6791 (from Stetson et al. 2003 photometry).
The upper right inset shows the position of the observed stars.}
\end{figure}

\begin{figure}
\includegraphics[scale=0.8]{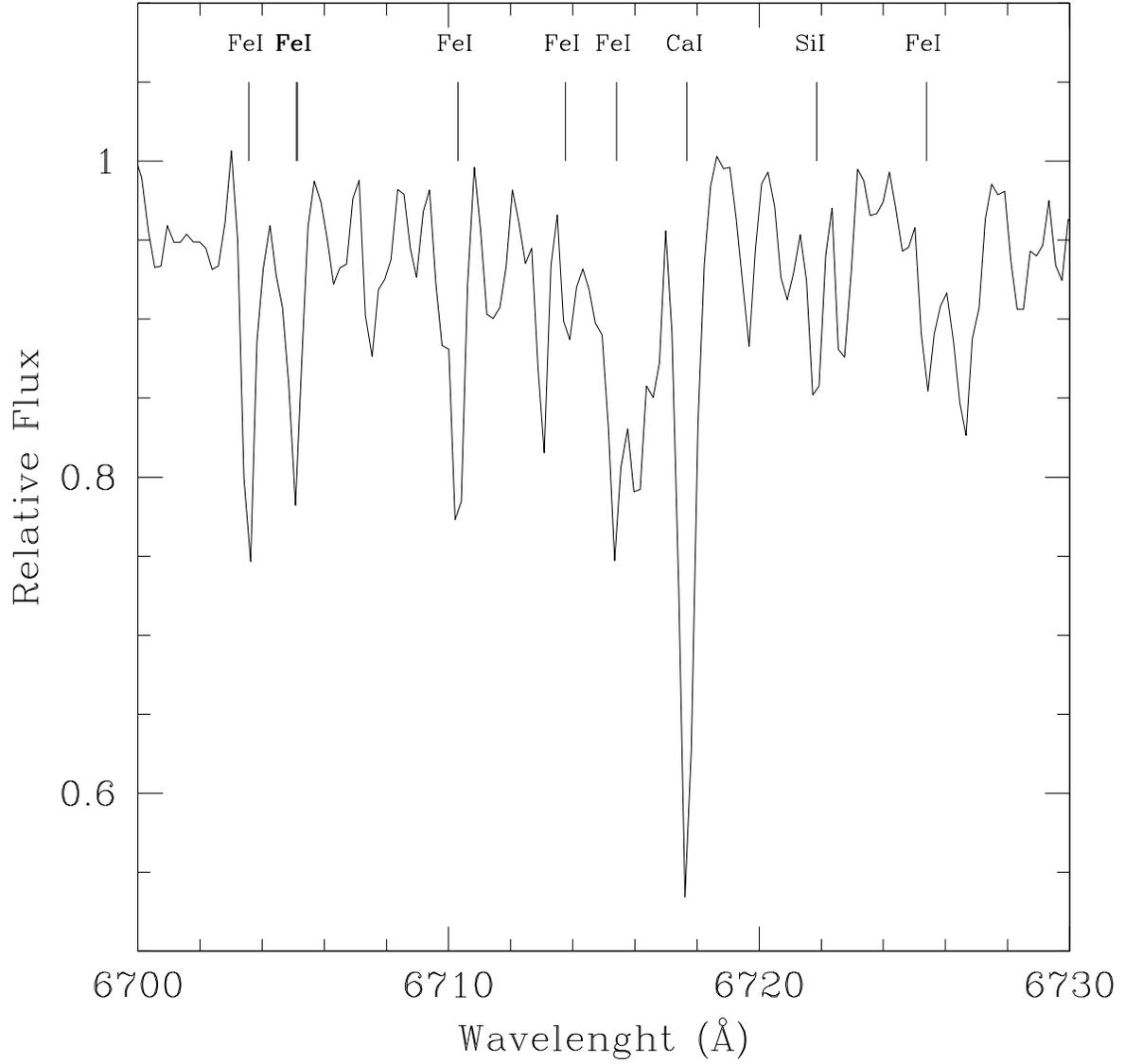}
\caption{An example of extracted spectrum for the star $\#$1181,
with the main lines indicated.}
\end{figure}

\begin{figure}
\includegraphics[scale=0.8]{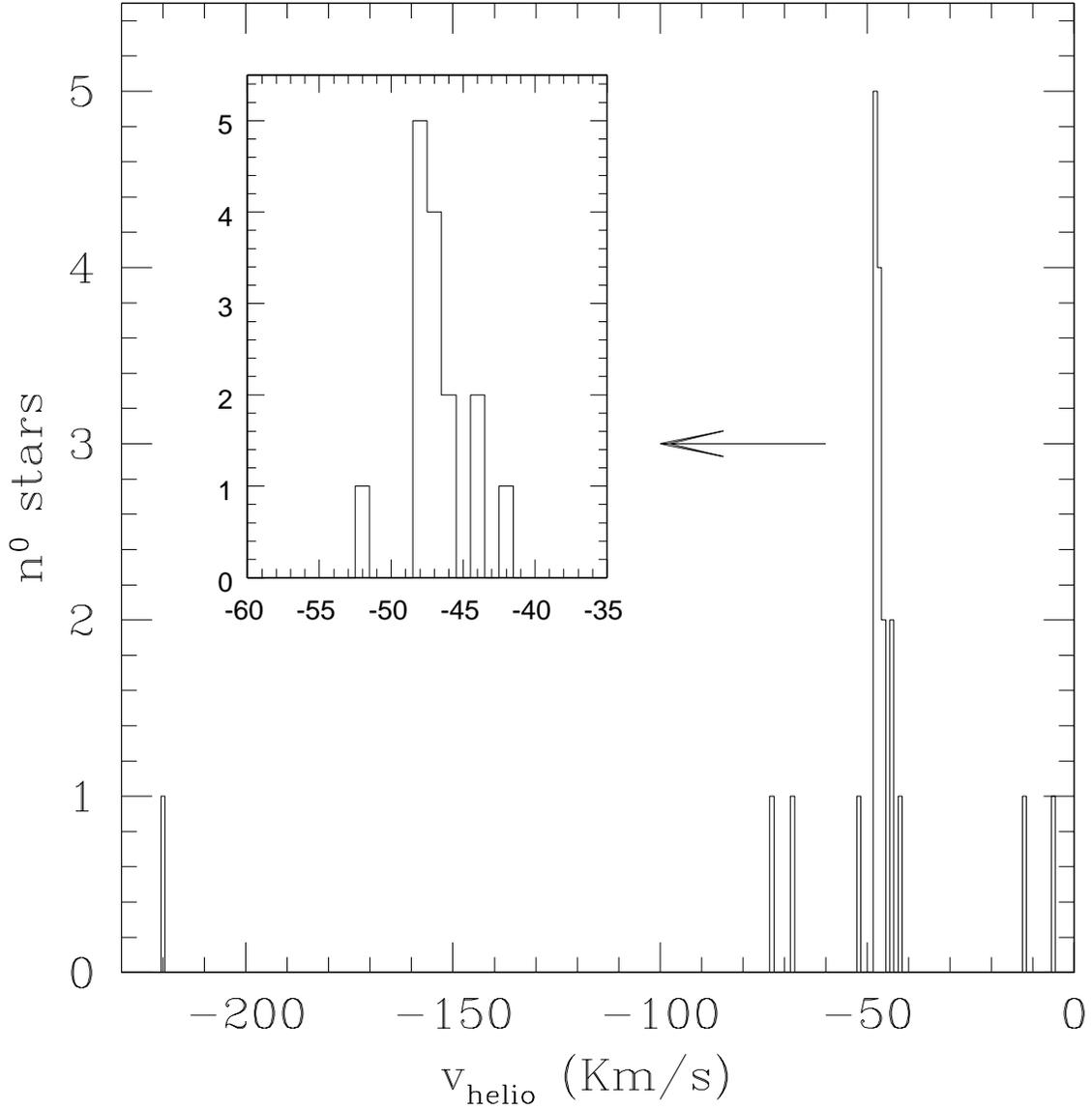}
\caption{Radial velocity distribution
of the 20 observed target stars.}
\end{figure}

\begin{figure}
\includegraphics[scale=0.8]{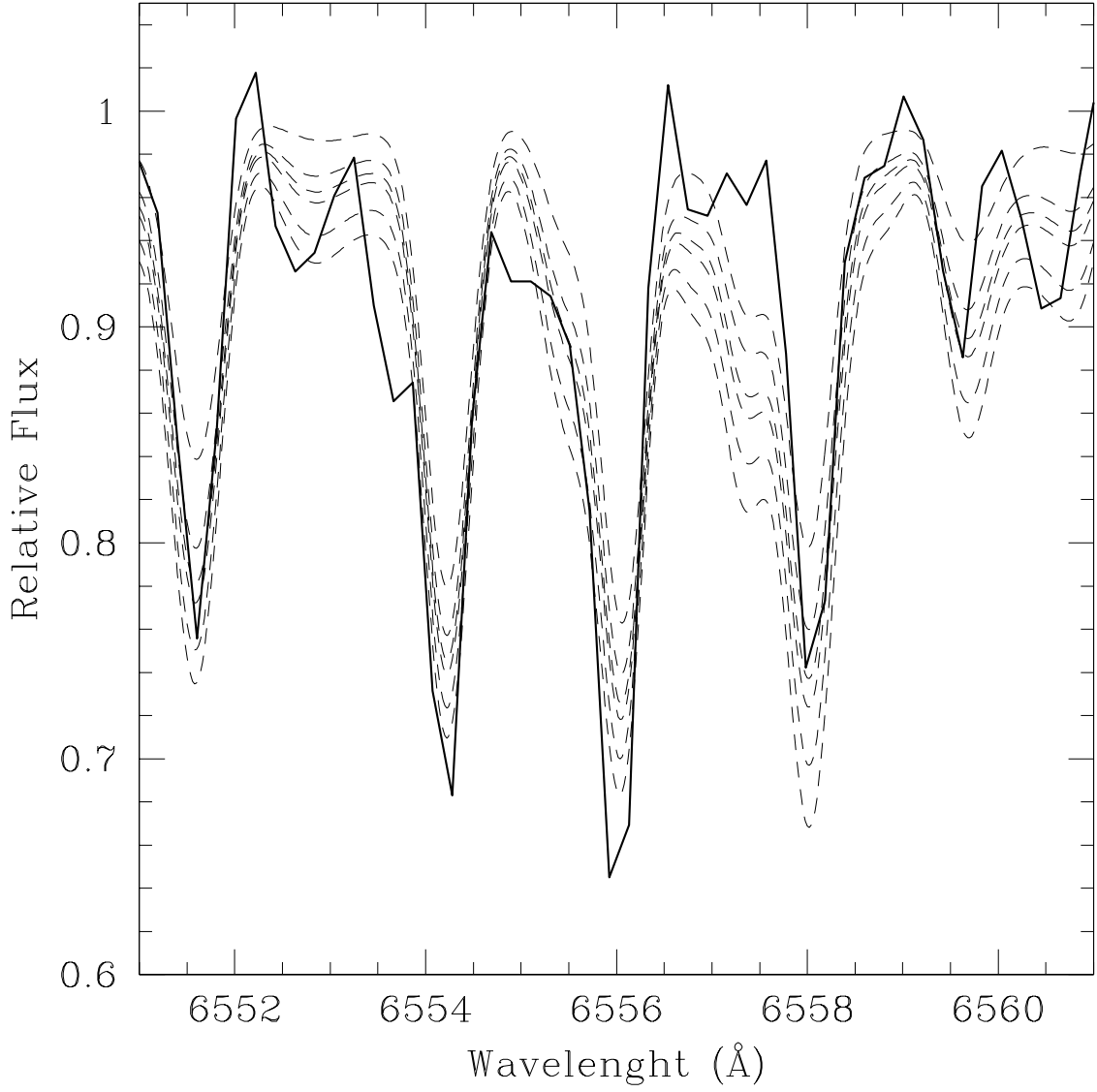}
\caption{The spectrum of Fig~2. (thicker line) and a set of
synthetic spectra for [M/H]= -0.2, 0.0, +0.2, +0.5 and +0.7,
from the top to the bottom.}
\end{figure}

\begin{figure}
\includegraphics[scale=0.8]{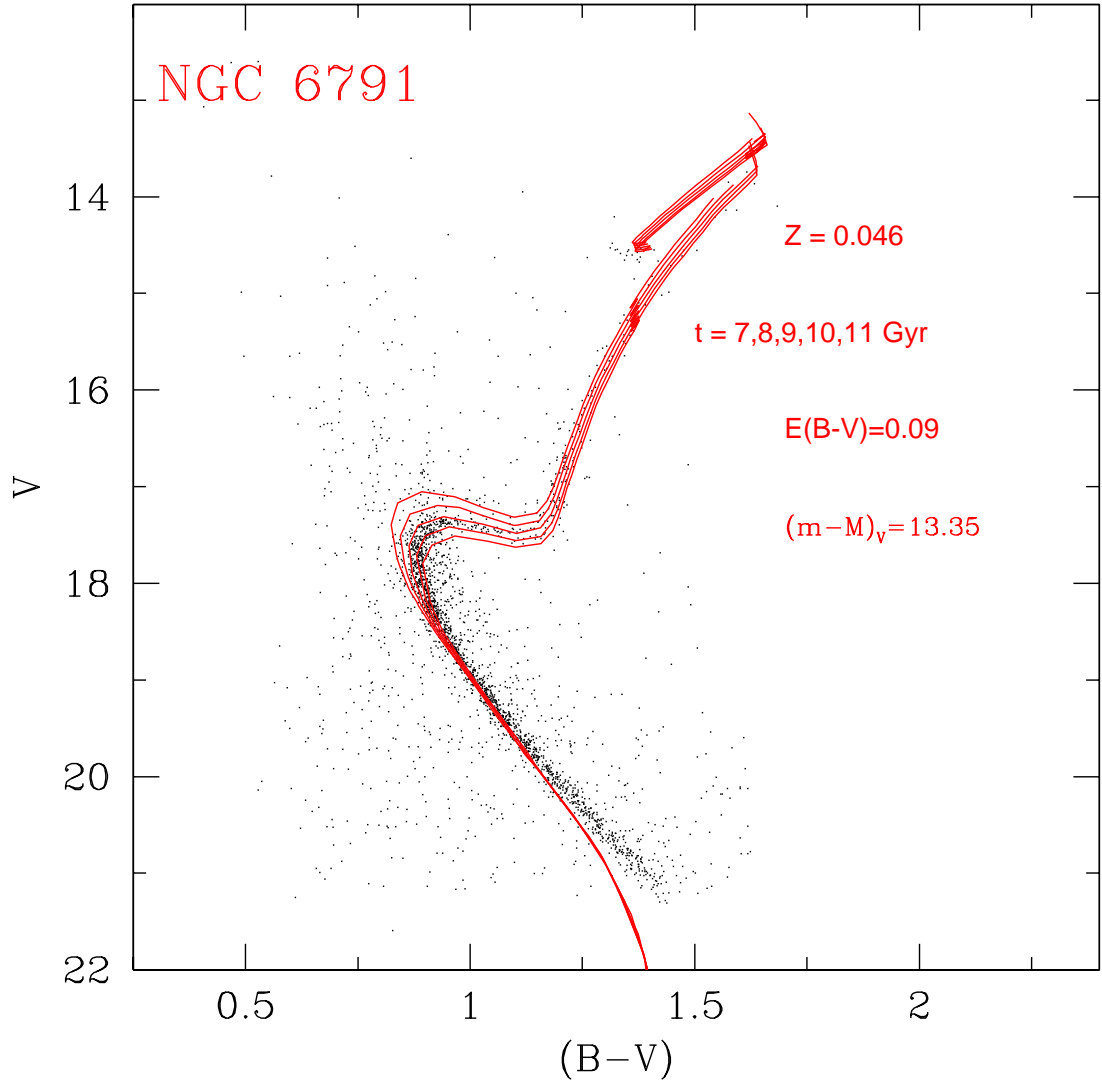}
\caption{
Five Padova isochrones for the ages of 7, 8, 9, 10, and 11
Gyr are fitted to the observed CMD 
in the V vs (B-V) plane. The
isochrones are shifted by $E(B-V)$ = 0.09 and $(m-M)_v$ = 13.35. 
}
\end{figure}

\begin{figure}
\includegraphics[scale=0.8]{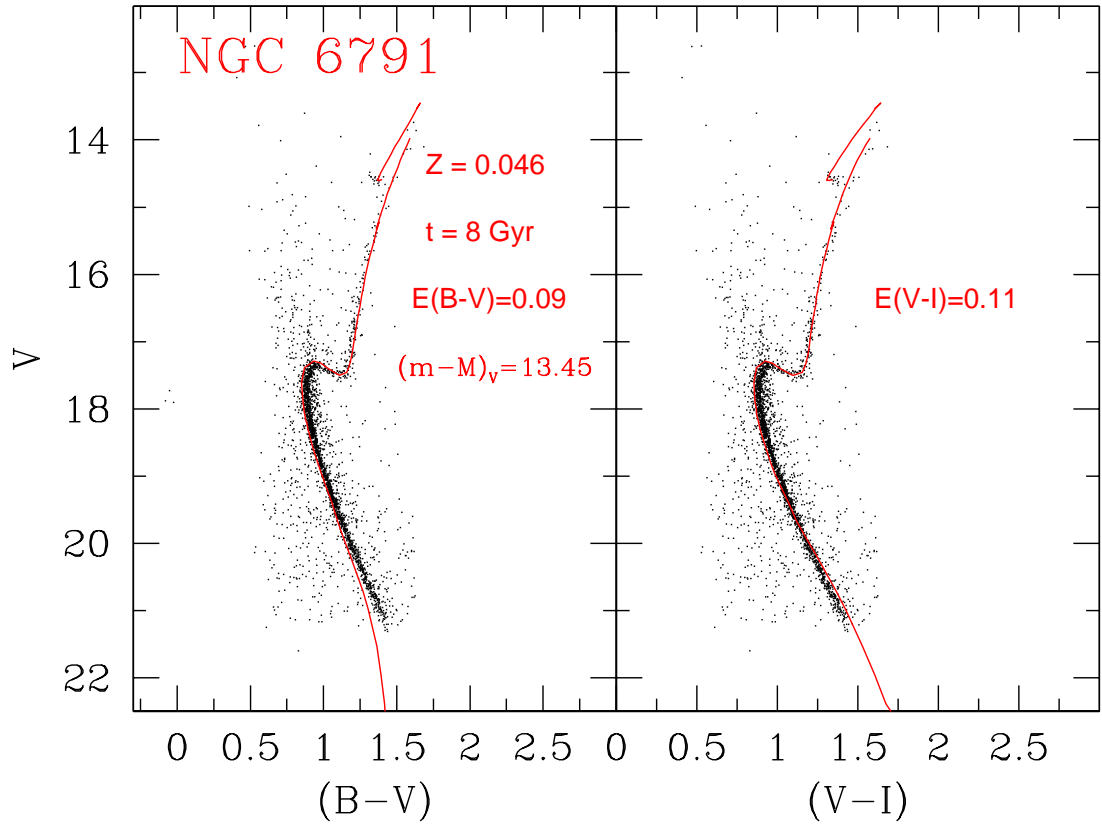}
\caption{Best fit isochrone solution of the CMD of NGC 6791 with 
the Padova models.
The isochrone and setting parameters are indicated in the plot.
}
\end{figure}

\begin{figure}
\includegraphics[scale=0.8]{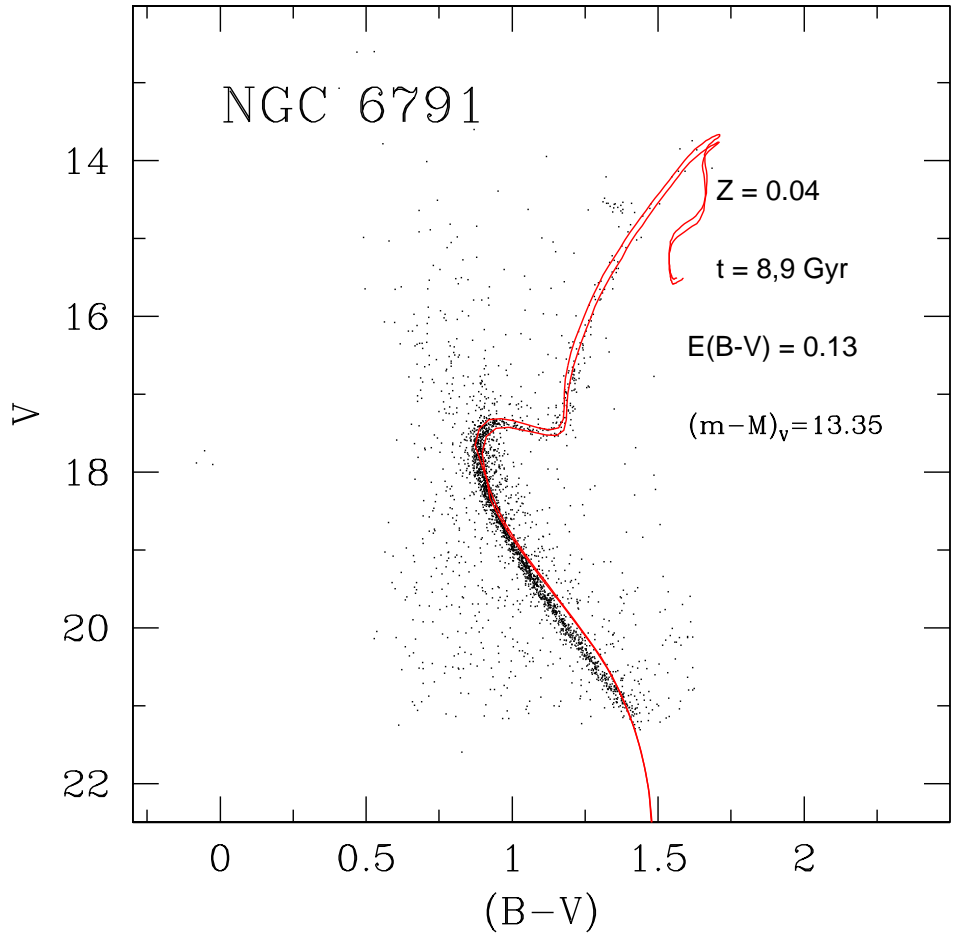}
\caption{Yale-Yonsei isochrone solution for ages of 8 and 9 Gyr. 
}
\end{figure}

\begin{figure}
\includegraphics[scale=0.8]{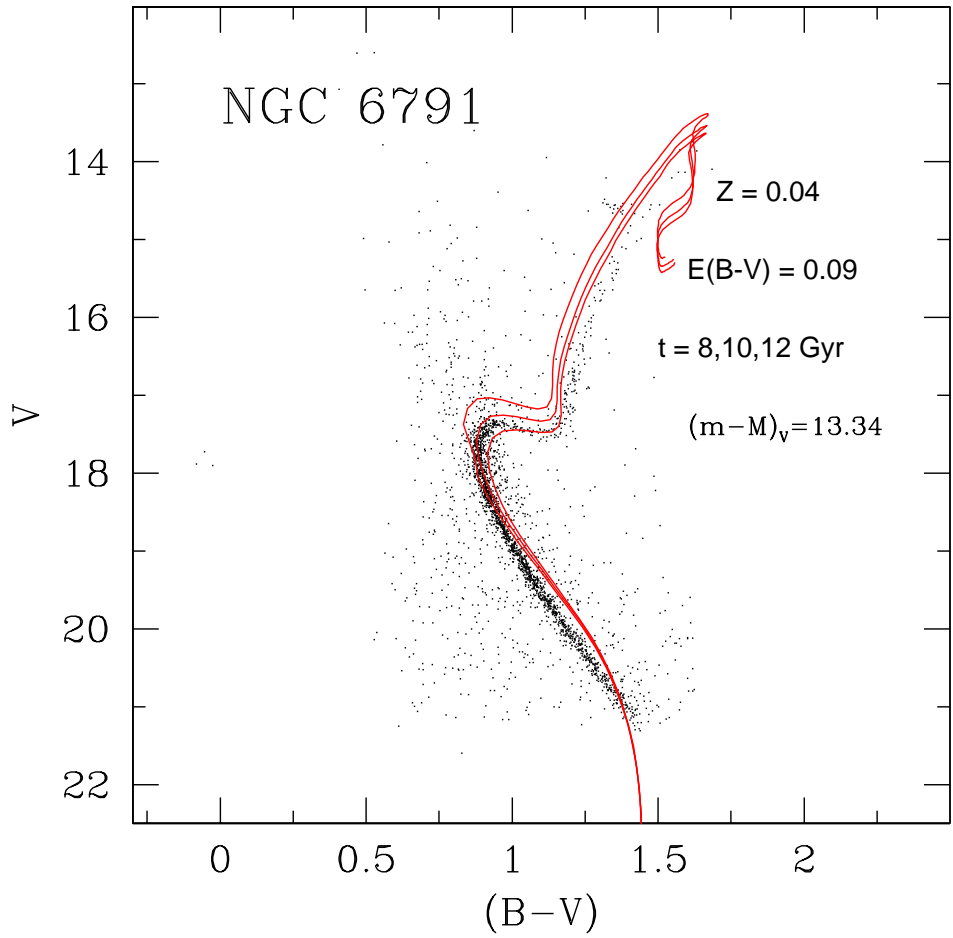}
\caption{Yale-Yonsei isochrone solution for ages of 8, 10 and 12 Gyr (from the top
to the bottom).
Note how also the two older age isochrones are clealy ruled out. 
}
\end{figure}

\begin{figure}
\includegraphics[scale=0.8]{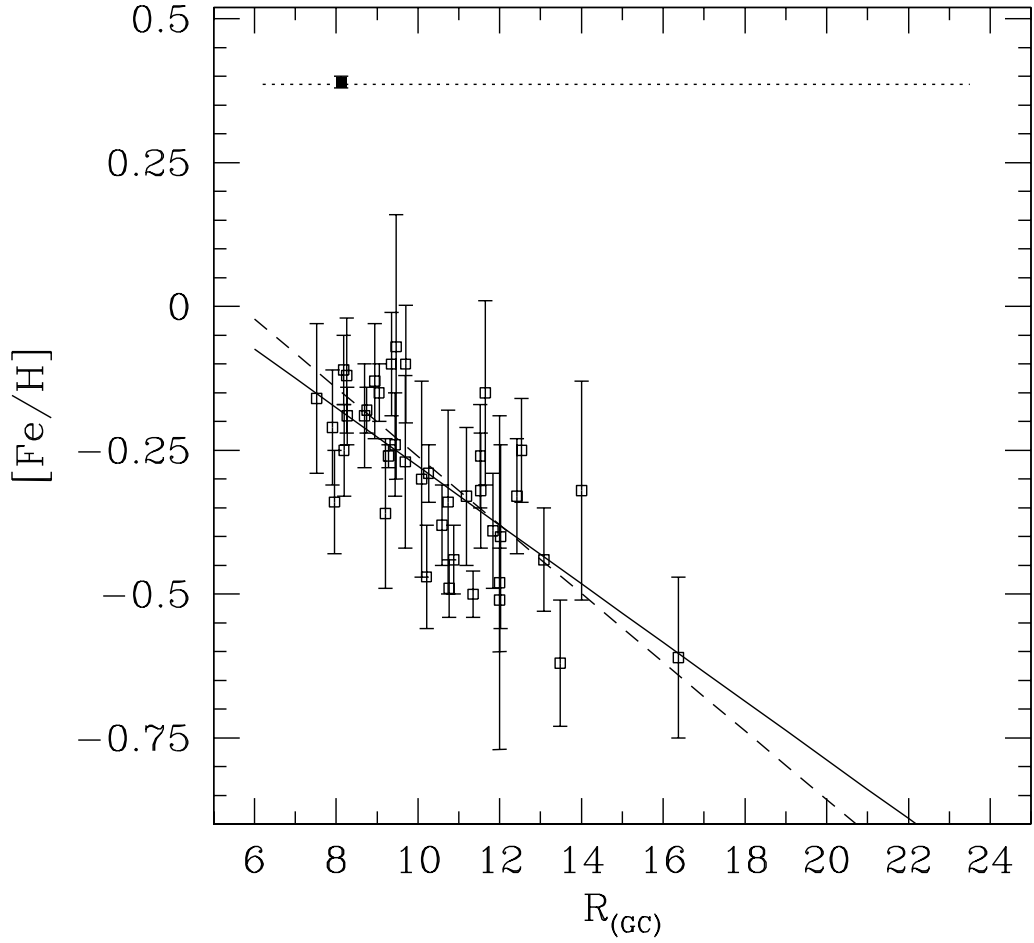}
\caption{The Galactic disk chemical abundance radial gradient.  The
data are from \citet{fri02}, with the exception of Berkeley~22 and
Berkeley~66 taken from \citet{vil05} and NGC~6791 (filled square, coming from the
present study). The solid line is the linear fit without NGC~6791,
whereas the dashed line is a linear fit to all the data points. The
horizontal dotted line shows the epicyclical amplitude of NGC 6791
orbit.}
\end{figure}

\begin{figure}
\plottwo{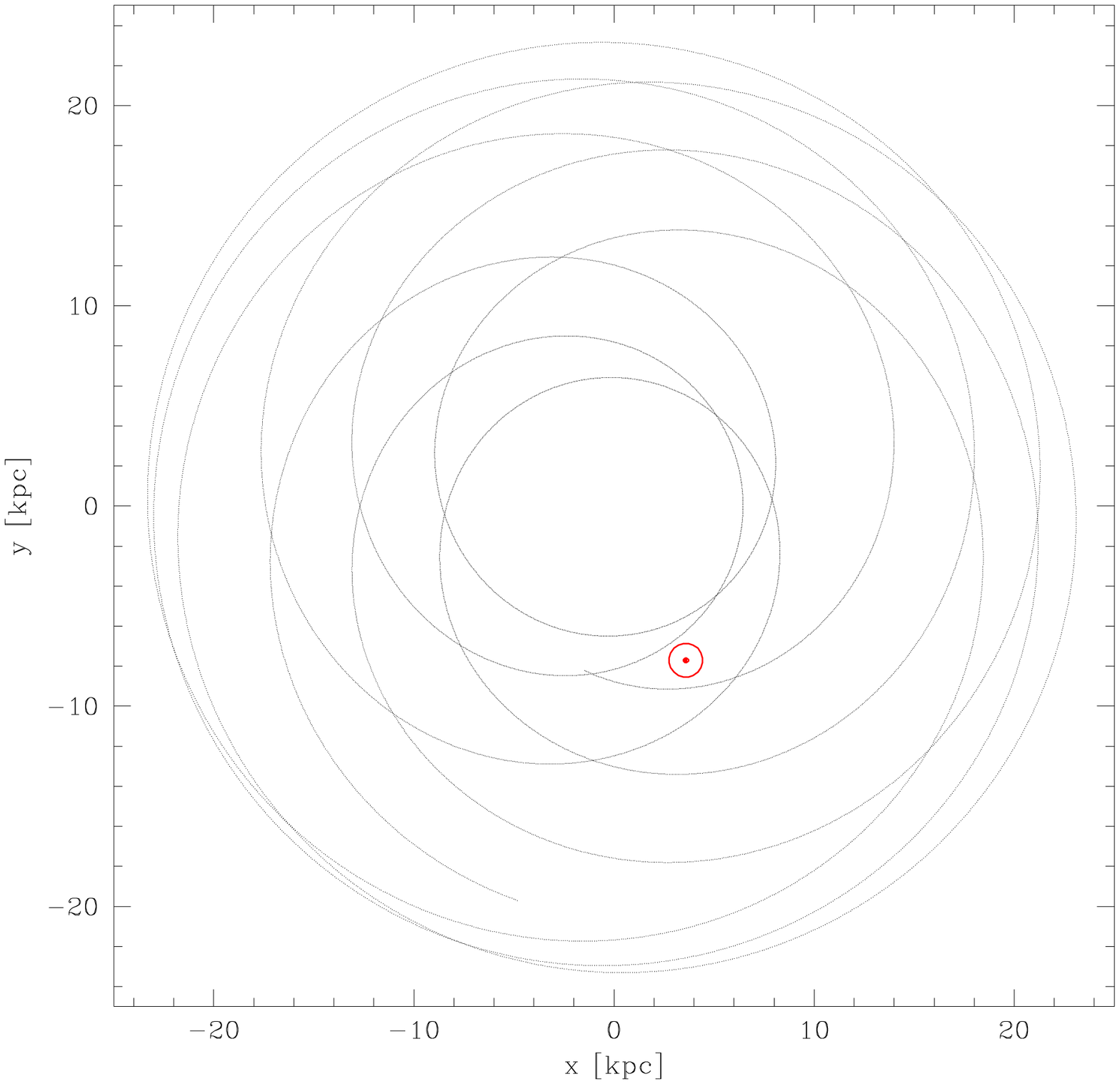}{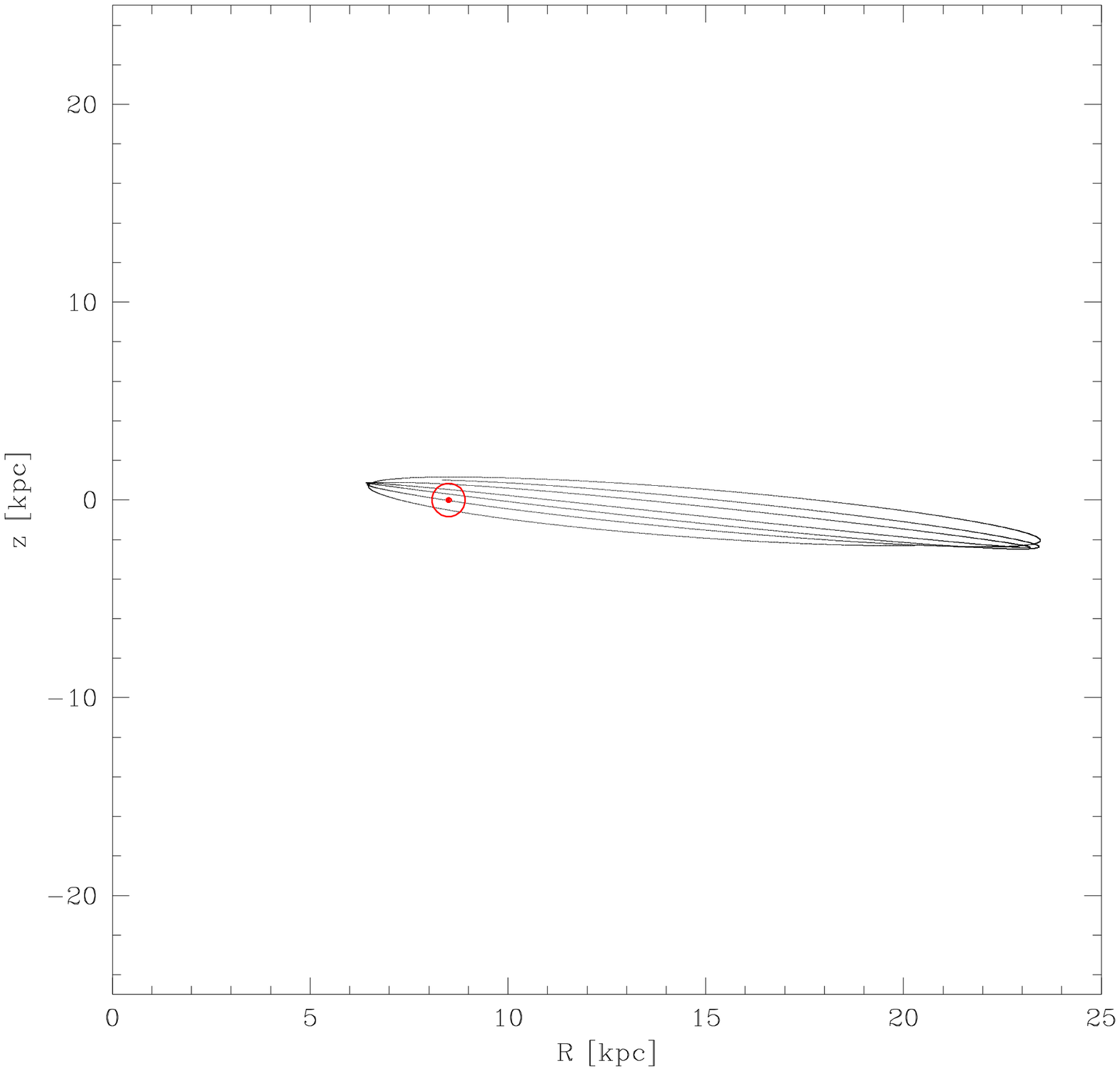}
\caption{ Galactic orbit of NGC 6791 in the X-Y and meridional plane.
The position of the Sun is indicated.}
\end{figure}

\clearpage

\begin{deluxetable}{lcccccccc}
\tabletypesize{\scriptsize}
\tablewidth{0pt}
\tablecaption{OBSERVED STARS}
\tablehead{
\colhead{ID} &  \colhead{K65} &  \colhead{RA} &  \colhead{DEC} &  \colhead{V} &  \colhead{(B-I)} &  \colhead{$V_{rad}$ (km s$^{-1}$)} &  \colhead{$S/N$} &  \colhead{Spectral Type}}
\startdata
10898  &        &  19:21:01.13 &  +37:42:13.80 &  14.459 &   3.059 &    -47.37$\pm$0.088 &   40 &  K4 III   \\
11814  &  3003  &  19:21:04.27 &  +37:47:18.90 &  13.849 &   3.233 &    -46.52$\pm$0.082 &   60 &  K4/5 III \\
1442   &        &  19:21:16.33 &  +37:52:15.80 &  14.056 &   2.580 &    -11.84$\pm$0.056 &  100 &           \\
2044   &        &  19:20:30.92 &  +37:48:45.30 &  14.144 &   2.917 &     -5.27$\pm$0.072 &   40 &           \\
2423   &        &  19:20:33.26 &  +37:50:12.70 &  14.082 &   2.794 &   -220.41$\pm$0.066 &   70 &           \\
2793   &  3036  &  19:20:34.86 &  +37:46:30.10 &  14.538 &   2.751 &    -48.00$\pm$0.108 &   30 &  M1/2 III \\
3369   &  3030  &  19:20:37.89 &  +37:44:49.30 &  14.529 &   2.673 &    -48.37$\pm$0.094 &   40 &  K2/3 III \\
4162   &        &  19:20:40.85 &  +37:46:21.80 &  14.551 &   2.687 &    -52.11$\pm$0.146 &   10 &  K2/3 III \\
4715   &        &  19:20:42.73 &  +37:51:07.70 &  14.515 &   2.698 &    -47.91$\pm$0.088 &   50 &  K2/3 III \\
5583   &        &  19:20:45.58 &  +37:39:51.20 &  14.602 &   2.742 &    -43.62$\pm$0.196 &   10 &  K3/4 III \\
6940   &  3013  &  19:20:49.67 &  +37:44:08.00 &  14.588 &   2.659 &    -46.00$\pm$0.155 &   10 &  K2/3 III \\
7922   &        &  19:20:52.47 &  +37:50:15.80 &  14.482 &   2.671 &    -48.28$\pm$0.080 &   30 &  K2/3 III \\
7972   &  3010  &  19:20:52.60 &  +37:44:28.50 &  14.136 &   3.356 &    -44.35$\pm$0.094 &   40 &  K7 III   \\
8082   &  SE-49 &  19:20:52.89 &  +37:45:33.40 &  14.546 &   2.639 &    -46.18$\pm$0.102 &   40 &  K2/3 III \\
8266   &  2001  &  19:20:53.39 &  +37:48:28.40 &  13.741 &   3.395 &    -47.73$\pm$0.080 &   40 &  K9 III   \\
852    &        &  19:20:22.40 &  +37:51:42.40 &  14.738 &   2.748 &    -67.91$\pm$0.267 &   10 &           \\
8563   &        &  19:20:54.19 &  +37:46:28.80 &  14.554 &   3.071 &    -42.44$\pm$0.084 &   40 &  K4 III   \\
8904   &  2008  &  19:20:55.11 &  +37:47:16.50 &  13.862 &   3.603 &    -46.91$\pm$0.098 &   30 &  M0 III   \\
8988   &  3018  &  19:20:55.31 &  +37:43:15.60 &  14.557 &   3.005 &    -47.46$\pm$0.088 &   30 &  K4 III   \\
95     &        &  19:20:11.19 &  +37:49:48.70 &  13.589 &   2.986 &    -72.93$\pm$0.076 &   50 &           \\
\enddata
\end{deluxetable}

\clearpage

\begin{deluxetable}{lccc}
\tabletypesize{\scriptsize}
\tablewidth{0pt}
\tablecaption{ADOPTED ATMOSPHERIC PARAMETERS}
\tablehead{
\colhead{ID} &  \colhead{T$_{eff}$ ($^oK$)} &  \colhead{log {\it g} (dex)} &  \colhead{$v_t$ (km s$^{-1}$)} }
\startdata
10898  &    4100  &   2.53 &    1.0 \\
11814  &    3980  &   2.17 &    1.1 \\
2793   &    3760  &   2.02 &    1.1 \\
3369   &    4400  &   2.79 &    1.0 \\
4162   &    4370  &   2.78 &    1.0 \\
4715   &    4360  &   2.76 &    1.0 \\
5583   &    4300  &   2.75 &    1.0 \\
6940   &    4400  &   2.81 &    1.0 \\
7922   &    4390  &   2.77 &    1.0 \\
7972   &    3920  &   2.23 &    1.1 \\
8082   &    4410  &   2.81 &    1.0 \\
8266   &    3900  &   2.04 &    1.2 \\
8563   &    4080  &   2.55 &    1.0 \\
8904   &    3830  &   2.01 &    1.2 \\
8988   &    4130  &   2.60 &    1.0 \\
\enddata
\end{deluxetable}

\clearpage

\begin{deluxetable}{lcccc}
\tabletypesize{\scriptsize}
\tablewidth{0pt}
\tablecaption{LINELIST}
\tablehead{}
\startdata
6408.016\ FeI &  6411.650\ FeI &  6419.980\ FeI &  6421.350\ FeI  &  6436.430\ FeI\\ 
6439.075\ CaI &  6449.808\ CaI &  6455.598\ CaI &  &6464.661\ FeI \\ 
6469.123\ FeI &  6469.210\ FeI &  6471.662\ CaI &  &6481.880\ FeI\\ 
6491.561\ Ti2 &  6494.980\ FeI &  6496.897\ BaII&  &6498.950\ FeI\\ 
6499.650\ CaI &  6501.691\ TiI &  6518.380\ FeI &  6527.202\ SiI  &NiI\\ 
6554.230\ TiI &  6556.070\ TiI &  6569.230\ FeI &  6572.779\ CaI  &  6574.240\ FeI\\ 
6575.020\ FeI &  6581.221\ FeI &  6593.880\ FeI &  6606.970\ TiII &  6608.030\ FeI\\ 
6609.120\ FeI &  6625.041\ FeI &  6627.560\ FeI &  6633.440\ FeI  &  6633.760\ FeI\\ 
6634.100\ FeI &  6643.640\ NiI &  6646.980\ FeI &  6663.231\ FeI  &  6663.450\ FeI\\ 
6677.955\ FeI &  6677.990\ FeI &  6698.673\ AlI &  6703.570\ FeI  &  6705.101\ FeI\\ 
6705.131\ FeI &  6710.310\ FeI &  6713.760\ FeI &  6715.410\ FeI  &  6717.681\ CaI\\ 
6721.848\ SiI &  6725.390\ FeI &  6733.160\ FeI &  6737.980\ FeI  &  6741.628\ SiI\\ 
6743.120\ TiI &  6743.185\ TiI &  6750.150\ FeI &                &              \\
\enddata
\end{deluxetable}

\clearpage

\begin{deluxetable}{lcccccccc}
\tabletypesize{\scriptsize}
\tablewidth{0pt}
\tablecaption{MEAN STELLAR ABUNDANCES}
\tablehead{
\colhead{ID} &  \colhead{[M/H]} &  \colhead{[Fe/H]} &  \colhead{[CaI/H]} &  \colhead{[Ti/H]} &   \colhead{[Ba/H]} &  \colhead{[Si/H]} &  \colhead{[Ni/H]} &  \colhead{[Al/H]} }
\startdata
10898 &   0.37$\pm$0.06 &  0.38$\pm$0.08 &  0.33$\pm$0.12 &  0.38$\pm$0.02 &         &   0.42          &   0.32         &   0.20\\
11814 &   0.39$\pm$0.09 &  0.34$\pm$0.08 &  0.32$\pm$0.08 &  0.34$\pm$0.05 &         &   0.39$\pm$0.03 &   0.29$\pm$0.14&   0.23\\
3369  &   0.41$\pm$0.09 &  0.37$\pm$0.09 &  0.35$\pm$0.05 &  0.38$\pm$0.07 &  0.19   &   0.42          &   0.36         &   0.18\\
4715  &   0.39$\pm$0.07 &  0.36$\pm$0.05 &  0.33$\pm$0.07 &  0.35$\pm$0.02 &  0.18   &   0.41$\pm$0.02 &   0.42         &   0.23\\
7922  &   0.39$\pm$0.02 &  0.39$\pm$0.07 &  0.33$\pm$0.14 &  0.36$\pm$0.02 &         &   0.37$\pm$0.09 &   0.40         &   0.22\\
7972  &   0.40$\pm$0.09 &  0.39$\pm$0.05 &  0.35$\pm$0.06 &  0.34$\pm$0.05 &  0.20   &   0.39          &   0.41         &   0.26\\
8082  &   0.42$\pm$0.04 &  0.38$\pm$0.05 &  0.38$\pm$0.07 &  0.38$\pm$0.07 &  0.14   &   0.36          &   0.48         &   0.23\\
8266  &   0.37$\pm$0.05 &  0.37$\pm$0.05 &  0.38$\pm$0.07 &  0.37          &  0.24   &   0.39          &   0.31         &   0.19\\
8563  &   0.38$\pm$0.08 &  0.38$\pm$0.06 &  0.32$\pm$0.03 &  0.36$\pm$0.04 &  0.32   &   0.42          &   0.37         &   0.25\\
8988  &   0.36$\pm$0.06 &  0.40$\pm$0.07 &  0.35$\pm$0.10 &  0.31$\pm$0.07 &  0.22   &   0.39          &   0.34         &       \\
\enddata
\end{deluxetable}

\begin{deluxetable}{lccccccc}
\tabletypesize{\scriptsize}
\tablewidth{0pt}
\tablecaption{ABUNDANCE RATIOS}
\tablehead{
\colhead{ID} &  \colhead{[Fe/H]} &   \colhead{[Ca/Fe]} &  \colhead{[Ti/Fe]} &  \colhead{[Ba/Fe]} &  \colhead{[Si/Fe]} &  \colhead{[Ni/Fe]} 
&  \colhead{[Al/Fe]}}
\startdata
10898 &   0.38$\pm$0.08 &  -0.05 &   0.00 &        &  +0.04 &  -0.06 &  -0.18\\
11814 &   0.34$\pm$0.08 &  -0.02 &   0.00 &        &  +0.05 &  -0.05 &  -0.11\\
3369  &   0.37$\pm$0.09 &  -0.02 &  +0.01 &  -0.14 &  +0.05 &  -0.01 &  -0.19\\
4715  &   0.36$\pm$0.05 &  -0.03 &  -0.01 &  -0.15 &  +0.05 &  +0.06 &  -0.13\\
7922  &   0.39$\pm$0.07 &  -0.06 &  -0.03 &        &  -0.02 &  +0.01 &  -0.17\\
7972  &   0.39$\pm$0.05 &  -0.04 &  -0.05 &  -0.15 &   0.00 &  +0.02 &  -0.13\\
8082  &   0.38$\pm$0.05 &   0.00 &   0.00 &  -0.22 &  -0.02 &  -0.02 &  -0.15\\
8266  &   0.37$\pm$0.05 &  +0.01 &   0.00 &  -0.12 &  +0.02 &  +0.02 &  -0.18\\
8563  &   0.38$\pm$0.06 &  -0.06 &  -0.06 &   0.00 &  +0.04 &  -0.01 &  -0.13\\
8988  &   0.40$\pm$0.07 &  -0.05 &  -0.06 &  -0.13 &  -0.01 &  -0.01 &       \\
\enddata
\end{deluxetable}

\end{document}